\documentclass[12pt,a4paper,twoside]{article}
\usepackage{epsfig}
\usepackage{baltlat1}
\usepackage{wrapfig}
\begin{document}
\ \
\vspace{0.5mm}

\setcounter{page}{1}
\vspace{8mm}

\titlehead{Baltic Astronomy, vol.12, XXX--XXX, 2003.}

\titleb{REDSHIFT MEASUREMENTS OF FAINT DISTANT\\GIANT 
RADIO GALAXIES AND OBSERVATIONAL\\ CONSTRAINTS ON THEIR 
JET POWER AND\\ DYNAMICAL AGE}

\begin{authorl}
\authorb{K.~Chy\.{z}y}{1} 
\authorb{M.~Jamrozy}{2}
\authorb{S.~J.~Kleinman}{3}
\authorb{J.~Krzesi\'{n}ski}{3}
\authorb{J.~Machalski}{1} 
\authorb{R.~McMillan}{3}
\authorb{A.~Nitta}{3}
\authorb{N.~Serafimovich}{4} and
\authorb{S.~Zola}{1}
\end{authorl}

\begin{addressl}
\addressb{1}{Astronomical Observatory, Jagellonian University,
ul. Orla 171,\\ 30-244 Cracow, Poland}

\addressb{2}{Radio Astronomical Institute, Bonn University, 
Auf dem H\"ugel 71,\\ 53-121 Bonn, Germany }

\addressb{3}{Apache Point Observatory, Sunspot, NM, USA}

\addressb{4}{Special Astrophysical Observatory, Russian Academy of 
Science, Russia}

\end{addressl}

\submitb{Received: xx October 2003}

\begin{abstract}
The redshifts of faint radio galaxies identified with {\sl giant} radio 
source candidates selected from the sample of Machalski et al. (2001) 
have been measured. Given the redshift, the projected linear size and 
radio luminosity are then determined. The above, supplemented with the 
axial ratio of the sources (evaluated from the radio maps) allows to 
constrain their jet power and the dynamical age using the analytical 
model of Kaiser et al. (1997) but modified by allowing the axial ratio 
of the source's cocoon to evolve in time.
\end{abstract}

%\begin{keywords}
%Galaxies: active -- Radio continuum: galaxies 
%\end{keywords}

\resthead{Redshift Measurements of Distant Giant Radio Galaxies}
{K.~Chy\.{z}y M.~Jamrozy S.~J.~Kleinman at al.}

%{Institution}{Author(s)}

%\def\ninepoint{\def\rm{\fam0\ninerm} \textfont0=\ninerm}

\sectionb{1}{THE SAMPLE AND ITS OPTICAL SPECTROSCOPY}

We continue a supplement of the sample of faint {\sl giant} radio 
galaxy candidates of Machalski, Jamrozy \& Zola (2001) with further 
optical and radio data. The redshifts obtained for the sample host 
galaxies brighter than about 18.5 mag in the $R$ photometric band are 
given there.

The optical low-resolution spectra of a number of identified galaxies 
fainter than $R\approx 18.5$ mag are obtained with the APO 3.5m 
telescope equipped with the two-side spectrograph DIS-II and covering 
the spectral range of 3750--5600 \AA (blue side) and 5500--9000 \AA 
(red side). A 1.5$^{\prime\prime}$ wide slit providing a dispersion
of 3.15~\AA\, per pixel and spectral resolution of about 7~\AA\, were 
used. The wavelength calibration was carried out using exposures to 
argon/neon/helium lamp, and the flux calibration by short exposures of 
a spectrophotometric standard star close to the observed galaxy. The 
limited exposures of about 30 min were taken which was sufficient to 
determine a redshift for galaxies with emission lines detected but not 
for galaxies with continuum emission and absorption bands only. 
Therefore, we have been able to determine redshifts for the radio 
galaxies J1604+3438, J1649+3114, J1712+3558 and J1725+3923. 
Preliminary spectra for J1330+3850 and J1513+3841 show no emission 
lines and their crude redshift is estimated from the shape of 
calibrated continuum compared by eye with the template spectrum of 
elliptical galaxies (e.g. Kennicutt 1992). Examples of the APO spectra 
are shown in Fig.~1a,\,b.

\begin{figure}[h]
\centerline{\psfig{figure=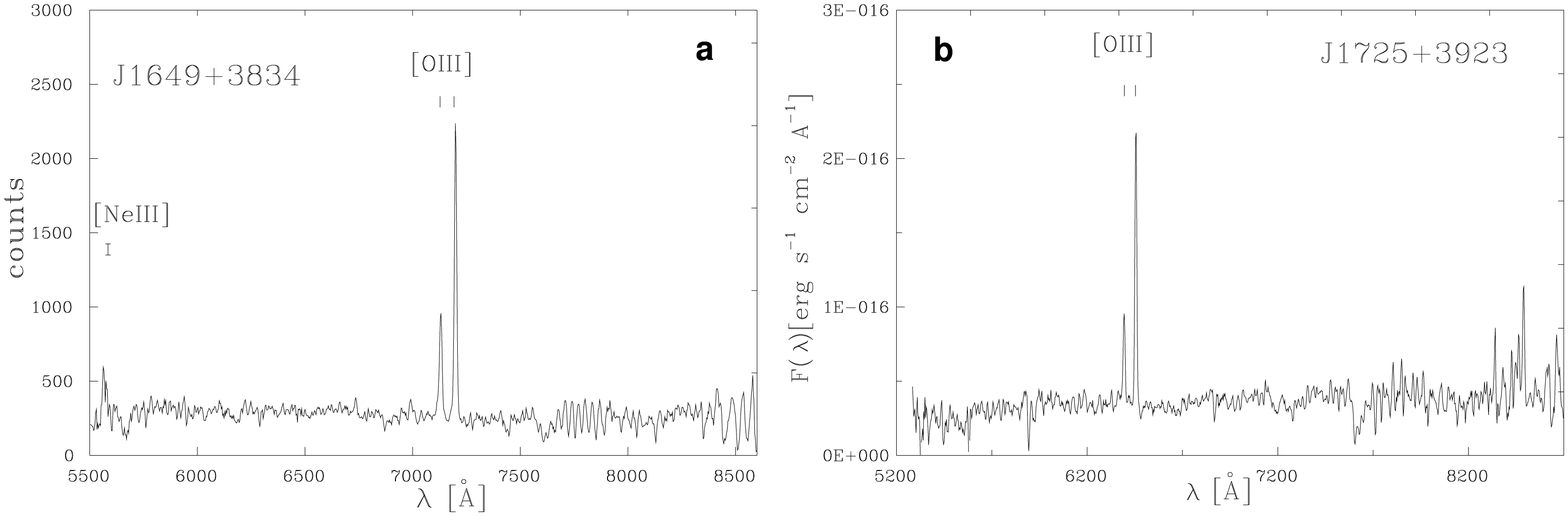,width=125truemm,angle=0,clip=}}
\captionb{1}{{\footnotesize Examples of the optical spectra of the observed galaxies}}
\end{figure}

\sectionb{3}{JET POWER, DYNAMICAL AGE, AND OTHER PARAMETERS}

For an observational constraint on physical parameters of the 
investigated faint {\sl giant} sources we use results of the Machalski 
et al.'s (2003) analysis. Their analysis is based on the analytical 
model of Kaiser et al. (1997) [hereafter KDA model] allowing to fit
these parameters for a source with given radio luminosity, size, 
volume, and age. Having the redshift {\footnotesize$z$}, linear size
{\footnotesize$D$}, 1.4 GHz luminosity, and axial ratio 
{\footnotesize$AR$} -- we fit the jet power {\footnotesize$Q_{jet}$} 
and density of the central core {\footnotesize$\rho_{0}$}
for a given source at different ages. A range of possible values of
{\footnotesize$Q_{jet}$} and the dynamical age of a source is limited 
by the range of {\footnotesize$\rho_{0}$}. The expected values of 
{\footnotesize$Q_{jet}$} and $\rho_{0}$ for different ages of the
investigated {\sl giants} are plotted in Fig.~2a,\,b,\,c.

Given {\footnotesize$Q_{jet}$} and $\rho_{0}$ enable one to calculate 
energy density ({\footnotesize$u_{c}$}) and pressure 
({\footnotesize$p_{c}$}) in the cocoon, total source energy 
({\footnotesize$E_{tot}$}), ratio of adiabatic to radiation losses, 
etc. given in Table~1.

\begin{figure}[t]
\centerline{{\psfig{figure=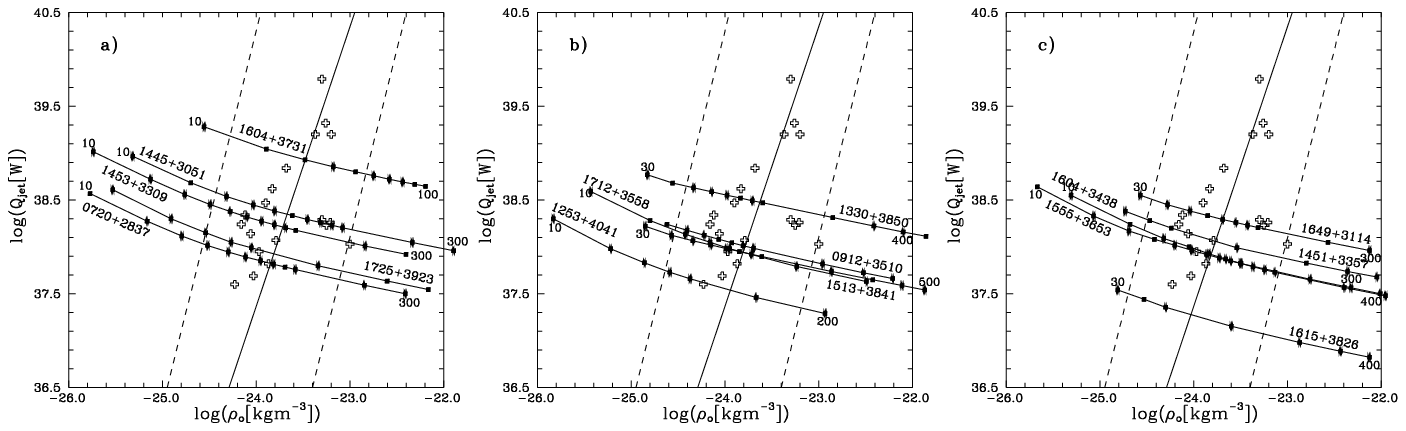,width=130truemm,angle=0,clip=}}}
\captionb{2}{{\footnotesize{\bf a, b, c} The range of possible values 
of {\footnotesize$Q_{jet}$} and dynamical age {\footnotesize$t_{dyn}$} 
for 15 {\sl giant} galaxies from the sample. The heavy crosses are 
known {\sl giant} radio sources used to determine the statistics of 
{\footnotesize$\rho_{0}$}. The diagonal dashed lines indicate a 
statistical 0.95 C.I. for this parameter. The diagonal solid line 
gives the most likely values of {\footnotesize$\rho_{0}$} adopted for 
the calculation of other physical parameters of the investigated 
sources. The age line for each named source begins and ends at the 
marked age values given in Myr}}
\end{figure}

\begin{center}
\vbox{\footnotesize
{\smallbf\ \ Table 1.}{\small\ Physical parameters of the investigated 
Giant radio galaxies derived from the observational data and expected 
from the model}\\
}
\end{center}
\begin{center}
\vbox{\footnotesize
\begin{tabular}{llcccccc}
%\hspace*{520mm}
Source      & $z$ &  $D$  &$\Delta$t(0.95 C.I.) & lg\,$Q_{jet}$ &lg\,$\rho_{0}$ & $E_{tot}$
& $\frac{2Q_{jet}t_{est}}{E_{c}}$\\
           &         & [Mpc]    & [Myr] & [W]      &[kg\,m$^{-3}$]& [J] \\
           &            &         &       &          &       &    \\
J0720+2837 & 0.2705 & 1.91 & 38--103 & 37.82 & $-$23.84 & 52.61 & 8.1\\
J0912+3510 & 0.2489 & 1.84 & 45--180 & 38.01 & $-$23.80 & 53.01 & 5.8\\
J1253+4041 & 0.2302 & 1.28 & 38--165 & 37.56 & $-$23.89 & 52.45 & 5.8\\
J1330+3850 & (0.63) & 2.70 & 47--185 & 38.48 & $-$23.64 & 53.43 & 6.8\\
J1445+3051 & 0.420  & 1.90 & 27--110 & 38.35 & $-$23.67 & 52.97 & 8.6\\
J1451+3357 & 0.3251 & 1.41 & 39--155 & 38.07 & $-$23.82 & 53.07 & 4.9\\
J1453+3309 & 0.249  & 1.57 & 41--170 & 38.21 & $-$23.71 & 53.28 & 4.7\\
J1513+3841 & (0.50) & 1.49 & 41--160 & 37.94 & $-$23.81 & 52.94 & 5.2\\
J1555+3653 & 0.2472 & 1.63 & 35--150 & 37.91 & $-$23.82 & 52.75 & 6.7\\
J1604+3438 & 0.2817 & 1.06 & 24--105 & 37.92 & $-$23.83 & 52.68 & 5.7\\
J1604+3731 & 0.814  & 1.50 & 13---54 & 38.93 & $-$23.48 & 53.24 & 9.3\\
J1615+3826 & 0.1853 & 1.10 & 34--140 & 37.30 & $-$23.96 & 52.08 & 6.4\\
J1649+3114 & 0.4373 & 1.40 & 33--135 & 38.29 & $-$23.69 & 53.25 & 4.9\\
J1712+3558 & 0.3357 & 1.23 & 27--115 & 37.96 & $-$23.77 & 52.72 & 6.0\\
J1725+2923 & 0.2898 & 1.45 & 30--130 & 37.94 & $-$23.70 & 52.70 & 7.1

\end{tabular}}
\end{center}

\noindent
The data in Table~1 may suggest that 
a dispersion of all physical parameters of the investigated sources is 
rather low. This is caused by our assumption of the mean value of 
$\rho_{0}$ for a particular source (intersection of the solid lines in 
Fig.~2a,\,b,\,c) used to determine the other physical parameters. None 
the less, the jet power of our {\sl giants} is very likely between 
{\footnotesize$10^{38}$} and {\footnotesize$10^{39}$} W, but for an 
individual source the uncertainty of {\footnotesize$Q_{jet}$} is about 
10\% only! However, the uncertainty of its dynamical age may be about 
200\% (cf. column~4 in Table~1) because the dispersion of $\rho_{0}$ 
is large.

A comparison of the values of some parameters in Table~1 with the 
corresponding values for already known {\sl giant} radio galaxies as 
well as `normal-size' FRII-type sources are presented in Fig.~3a,\,b,\,c. 
Parameters of the investigated sources confirm that {\sl giants} 
achieve the lowest values of energy density and pressure in their 
cocoons.

\begin{figure}[t]
\centerline{{\psfig{figure=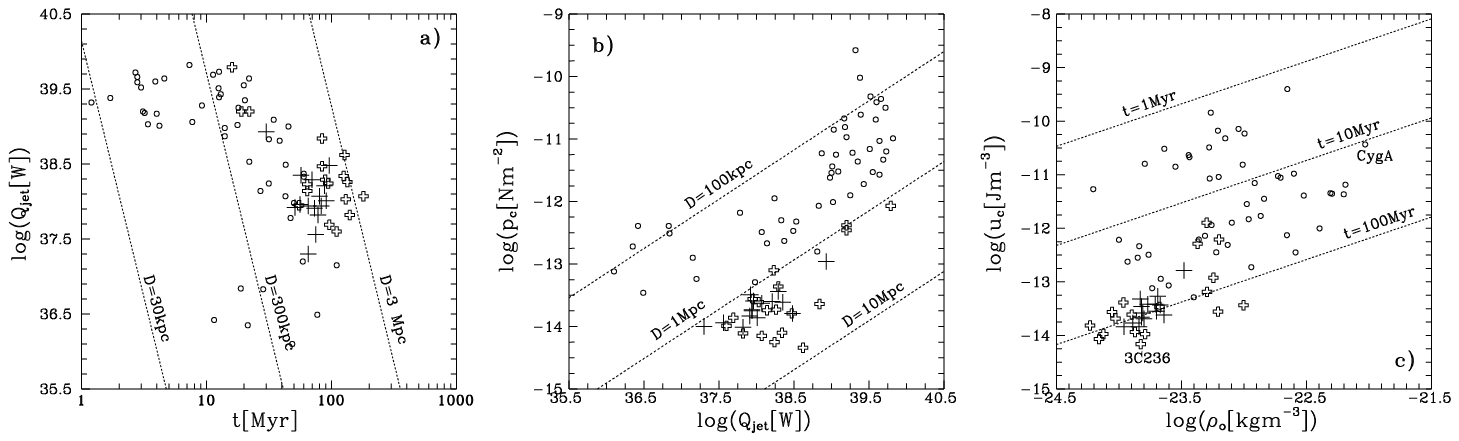,width=130truemm,angle=0,clip=}}}
\captionb{3}{{\footnotesize {\bf a)} Plot of the jet power 
{\footnotesize$Q_{jet}$} against source age {\footnotesize$t$}. 
The symbols on each of the three panels indicate: known Giants -- heavy 
crosses, the present Giants --large crosses, normal-size sources -- 
small circles. The dotted lines mark a constant linear size predicted 
from the statistical relation between {\footnotesize$D$},
{\footnotesize$Q_{jet}$}, and {\footnotesize$t$}. {\bf b)} The cocoon 
pressure {\footnotesize$p_{c}$} against {\footnotesize$Q_{jet}$}. The 
dotted lines mark a constant linear size of sources from the 
correlation between {\footnotesize$p_{c}$}, {\footnotesize$Q_{jet}$}, 
and {\footnotesize$D$}. {\bf c)} Energy density of the cocoon 
{\footnotesize$u_{c}$} against the central core density
{\footnotesize$\rho_{0}$}. The dotted lines show a constant age
from the statistical correlation between {\footnotesize$u_{c}$}, 
{\footnotesize$\rho_{0}$}, and {\footnotesize$t$}}}
\end{figure}

ACKNOWLEDGMENTS.\ MJ acknowledges the financial support from EAS.
\goodbreak

References\\
%\ref
Kaiser~C.~R., Dennett-Thorpe,~J., Alexander~P. 1997, MNRAS, 292, 723\\
%\ref
Kunnicutt~R.~C. 1992, ApJS, 79, 255\\
%\ref
Machalski~J., Jamrozy~M., Zola~S. 2001, A\&A, 371, 445\\
%\ref
Machalski~J., Chy\.{z}y~K.~T., Jamrozy~M. 2003, (submitted for MNRAS, 
astro-ph/0210546)

\end{document}